\pdfoutput=1
\newif\ifFull
\Fulltrue

\documentclass {journal}

\usepackage {a4wide}
\usepackage {url,hyperref}
\usepackage {amsmath}
\usepackage {amssymb}
\usepackage {plaatjes}
\usepackage {color}
\usepackage [pagewise]{lineno}

\newcommand {\eps} {\varepsilon}

\newcommand {\etal}{\textit{et~al.}}

\newcommand {\mathset} [1] {\ensuremath {\mathbb {#1}}}
\newcommand {\R} {\mathset {R}}

\newcommand {\N} {\mathset {N}}

\newtheorem {theorem} {Theorem}[section]
\newtheorem {problem}[theorem] {Problem}
\newtheorem {lemma}[theorem] {Lemma}

\newtheorem {corollary}[theorem] {Corollary}
\newtheorem {claim}[theorem] {Claim}

\newenvironment{proof}{\textbf {Proof:}}{\hfill $\boxtimes$}

\definecolor {infocolor} {rgb} {0.6,0.6,0.6}

\linenumbers

\definecolor {sepia} {rgb} {0.75,0.30,0.15}
\everymath{\color{sepia}}

\newcommand{\marrow}{\marginpar[\hfill$\longrightarrow$]{$\longleftarrow$}}
\newcommand{\beautifulremark}[3]{\textcolor{blue}{\textsc{#1 #2:}}
\textcolor{red}{\marrow\textsf{#3}}}
\renewcommand{\beautifulremark}[3]{} 

\newcommand{\maarten}[2][says]{\beautifulremark{Maarten}{#1}{#2}}

\newcommand{\joe}[2][says]{\beautifulremark{Joe}{#1}{#2}}

\newcommand{\reviewer}[2][says]{\beautifulremark{Reviewer}{#1}{#2}}

\title {Dynamic Planar Point Location\\ with Sub-Logarithmic Local Updates}
\shorttitle {Dynamic Planar Point Location with Sub-Logarithmic Local Updates}

\author
{
  Maarten L\"offler\thanks
  {
    Department of Computing and Information Sciences,
    Utrecht University;
    \texttt{m.loffler@uu.nl}.
  }
\and
  Joseph A. Simons\thanks
  {
    Computer Science Department,
    University of California, Irvine;
    \texttt{\{jsimons$|$dstrash\}@uci.edu}.
  }
\and
  Darren Strash\footnotemark[2]
}
\shortauthor{L\"offler~\etal}

\begin {document}
\maketitle
\begin {abstract}
  We study planar point location in a collection of disjoint fat regions, and investigate the complexity of \emph {local updates}:
  replacing any region by a different region that is ``similar'' to the original
  region. (i.e., the size differs by at most a constant factor, and distance
  between the two regions is a constant times that size). We show that it is
  possible to create a linear size data structure that allows for 
  insertions, deletions, and queries in logarithmic time, and allows for local updates
  in sub-logarithmic time on a pointer machine. 
  We begin by describing a solution for the $1$-dimensional version of the problem, where
  we can achieve constant time local updates.
  Then we show how the ideas can be extended to $2$ dimensions.

    We show that given constant similarity and fatness parameters:
    \begin {itemize}
      \item
    A set of $n$ disjoint intervals in $\R^1$ can be maintained in an $O(n)$
    size data structure that supports $O(\log n)$ worst-case time insertion,
    deletion, and point
    location queries, and $O(1)$ worst-case time local updates 
    (Section~\ref {sec:1d}).
    The data structure can be implemented on a
    real-valued pointer-machine.  
  \item A set of $n$ disjoint fat regions in $\R^2$ can be maintained
  in an $O(n)$ size data structure that supports $O(\log n)$ worst-case time
  insertion, deletion and
  point location queries, and $O(\log \log n)$ worst-case time local updates
  (Section~\ref {sec:2d}).
    The data structure can be implemented on a real-valued pointer-machine.

  \item We also give bounds that can handle arbitrary similarity and fatness
  parameters in Theorem~\ref{thm:1d-result} and Theorem~\ref{thm:2d-result} for
  the $\R^1$ and $\R^2$ case respectively.
 	\end {itemize} 

\end {abstract}


\thispagestyle{empty}
\setcounter{page}{0}
\clearpage

\section {Introduction}

  Planar point location lies at the heart of many geometric problems, and has
  been a major research topic in computational geometry for the past 40 years.
  In the static version of the problem, one aims to store a subdivision of the
  plane such that given a query point $q$ in the plane, the cell of the
  subdivision containing $q$ can be retrieved quickly~%
    \cite {dl-msp-76,egs-oplms-86,k-osps-83,lp-lppsi-77,st-pplpst-86}.
  In the dynamic
  version of the problem, one also allows changes to the data set,
  typically adding or removing line segments to the subdivision~%
    \cite {abg-idppl-06,b-skrp-77,gk-odvrs-09,gt-dtdpl-98,mn-dfc-90}.
  
  The best known dynamic data structures on a real RAM are due to Cheng and
  Janardan~\cite {cj-nrdpp-92}, who achieve $O(\log^2 n)$ queries and $O(\log
  n)$ updates where $n$ is the size of the subdivision, 
  and Arge \etal~\cite {abg-idppl-06}, who achieve $O(\log n)$
  queries, 
  $O(\log^{1+\eps}n)$ insertions, and $O(\log^{2+\eps}n)$ deletions.  A
  central open problem in this area is whether a linear-size data structure
  exists that can support both queries \emph {and} updates in logarithmic time,
  although this is known to be possible in more specific settings such as
  monotone or rectilinear subdivisions~%
    \cite{b-sedopl-08,gk-odvrs-09,gt-dtdpl-98}.
  Husfeldt \etal~\cite {hrs-lbdtcpplpm-96} prove that even in the very strong
  \emph {cell probe model}, there are $\Omega (\log n / \log \log n)$ lower
  bounds on both queries and updates.

Despite these theoretical results, practical evidence suggests that \emph {updating} a
  data structure should be fast.  Intuitively, an update to a data set should
  not need to depend on $n$ at all, unless we need to find the place where the
  update takes place (i.e., we need to do a point location query).  In this
  paper, we study point location data structures on a collection of fat
  objects in the plane
  that support \emph {local updates}: replace any region by a different region
  that is \emph{similar} 
  to the original.  
  We show that the lower bounds on
  updates can be broken in this setting, while still allowing $O(\log n)$
  queries and using $O(n)$ storage.

  The idea of local updates is not new. For example, Nekrich~\cite {y-dsluo-08} considers
  (on a word-RAM)
the local update operation insert$_\Delta(x,y)$ which inserts a new element $x$ into a $1$-dimensional sorted list,
given a pointer to an existing element $y$ that satisfies $|x - y| \leq \Delta$ for some distance parameter
$\Delta$. 
There is also a related concept called  \emph{finger updates}, where the position of the
update is known; see e.g.  
Fleischer~\cite{f-sbst-93}. 
 However, our results are the first in this area that work in a geometric
 setting, and they can be implemented on a real-valued pointer machine.
 (See Appendix~\ref {sec:compmodel} for a discussion of computation models.)

\ifFull
\joe{This paragraph doesn't add much compared to the space it takes}
In order to obtain our results, we develop several tools which we believe are interesting in their own right, such as a dynamic balanced compressed quadtree with worst-case constant time updates, and a tree decomposition that supports logarithmic searches and constant time local changes (see Section~\ref {sec:imp}).
\fi

  \subsection {Problem description}
 \label{sec:prob} 
  We define the problem in general dimension $d$, but restrict our attention to $d \in \{1,2\}$ in the remainder of this paper.
  %
	We use $|R|$ to denote the diameter of a region $R \subset \R^d$, that is, $|R| = \max_{p,q \in R} |pq|$.
  We say two \emph {fat}\footnote
  {We formally define fat regions in Section~\ref{sec:fatdef}.}
  regions $R_1, R_2 \subset \R^d$ are \emph
  {$\rho$-similar} if $|R_1 \cup R_2| \leq \rho \min\{|R_1|, |R_2|\}$, 
  see Figure~\ref {fig:intro-rhosimilar}.\footnote
{This definition captures two ideas at once: firstly, the \emph {sizes} of $R_1$ and $R_2$
can differ by at most a factor of $\rho$, and secondly, the
\emph {distance} between $R_1$ and $R_2$ can be at most a factor $\rho$ times the smaller
of these sizes.}
  
    \begin {problem}
      Given a set $\mathcal R$ of $n$ disjoint fat 
    regions in $\R^d$, store them in a data structure that allows:
    \begin {itemize}
        \item queries: given a point $q \in \R^d$, return the region in $\mathcal R$
        that contains $q$ (if any) in $Q(n)$ time;
        \item local updates: given a region $R \in \mathcal R$ and a region $R'$ that is 
    $\rho$-similar 
                to $R$, 
    replace $R$ by $R'$ in the data structure in $U(n)$ time; and
        \item global updates: delete an existing region $R$ from the data structure
    or insert a new region $R'$ into the data structure in $Q(n) + U(n)$
    time
    \end {itemize}
 such that $Q(n)=O(\log n)$ but $U(n)=o(\log n)$. Note that a local update
 allows for an arbitrary number of smaller regions to be ``between'' the old region $R$ and
 the new region $R'$. 
    \end {problem}
%
  \eenplaatje {intro-rhosimilar} 
  { Two $\rho$-similar regions for $\rho = \frac {d_{12}} {\min \{d_1,d_2\}}$.
  }
  \subsection {Applications}
  
\paragraph{Tracking moving objects.} 
A natural application of our data structure is to keep track of moving objects.
One may imagine a number of objects of different sizes moving unpredictably in an environment at different speeds.
A popular method 
for dealing with moving objects is to discretize time and 
process the new locations of the objects at each time step.
The naive way to do this is to simply
rebuild an entire data structure every time step.  
Our data structure can be used to process such changes more efficiently.

In computational geometry, there is a large literature on dealing with moving objects (or points).
\emph{Kinetic data structures}
are based on the premise that a data structure should not need to be updated each time step,
but rather only when some combinatorial feature of a description of objects changes~%
    \cite{aeg-kbisd-98,bgsz-pekds-97,ghsz-kcud-00,g-kdssar-98}.
A fundamental underlying assumption in kinetic data structures is that
trajectories of the moving objects are predictable, at least in the short term.
However, in many modern real-world
scenarios, trajectories are not predetermined, they are discovered in an online and inherently discrete fashion.
As a result, several theoretical approaches to deal with unpredictable motion
have been suggested recently, in various 
settings~%
    \cite {cmp-mnntim-09,brs-kchdtbbm-11,egl-tmofh-11,agnrz-cddn-04,mnpsw-cfim-04,yz-mdot-09}.
A common assumption in these works is to bound the maximum displacement after each update
(or velocity) of the 
moving points.
An interesting feature of our data structure is that we can simultaneously maintain objects moving at very different scales,
with a velocity bound that is dependent on the \emph {size} of the object.

\paragraph{Data imprecision.}
A different motivation for studying this problem comes from the desire to cope with
\emph {data imprecision}.  One way to model an imprecise point is to keep track
of a region of 
possible locations of the point%
    ~\cite {gss-cscah-93,nt-teb-00} (see also \cite{loffler-2009-phd} and the references therein).
 Recently, there has been a lot of
activity in this area~%
    \cite{cetal-cgur-10,djlz-lachaas-11,jlp-gcip-11,wz-mpchipcr-11}.
  Although algorithms to deal with imprecise data are beginning
to be well understood in a static setting, little effort has been devoted to
dealing with \emph {dynamic} imprecise points.  However, in many settings
imprecision is inherently dynamic (e.g. time-dependent or ``stale'' data), or
explicitly made dynamic (e.g. updates from new samples of the same point).

One of the simplest geometric queries on a data
structure that stores a point set one can imagine is the identity query. Given a query point, is
there a point in the data structure that is equal to the query point? When the
points in the data structure are imprecise, the answer to this question may have
three possible values: ``certainly'', ``possibly'', or ``certainly not.''
Distinguishing between the second and last answer\footnote {Under the mild assumption that
all points have at least some imprecision and there are only finitely many
points, the first answer will never occur.} comes down to testing whether
the query point (which we assume is a precise point) is contained in any of the
uncertainty regions of the imprecise points.  Therefore, we may view the problem
as a dynamic point location problem in a set of changing regions.

\ifFull
If we only wish to support increased precision updates (which would correspond
to stationary, but imprecise points), this question is closely related to
existing work in the update complexity
model~%
    \cite{bhkr-eusgcu-05,fggt-chapa-94},
in which one attempts to minimize the
number (or amount of gained precision) of updates necessary to correctly output
some structure; and to work on preprocessing imprecise 
points~%
    \cite{blmm-pipstse-10,Devillers10,hm-icg-08,ls-dtip-10,klm-pipst-10}, 
in which one
tries to prepare a set of imprecise points for faster computation of some
structure on the precise points once they become available. While these results
do not analyze the time complexity of single updates, they do provide some
evidence that sub-logarithmic update time may be possible.
\fi

  \subsection {Solution outline}
Geometric data structures are often either based on a recursive decomposition of the
data (e.g. a binary search tree) or a recursive decomposition of space (e.g. a
quadtree). 
Neither of those techniques by themselves are strong enough to solve the problem at hand,
so our solution combines both techniques.
We base our solution on a dynamic balanced\footnote{The quadtree is balanced in a
geometric sense, but may still have linear depth. See Section~\ref{sec:tools}} compressed
\emph{quadtree}~\cite{PreparataSh85},
the details of which are covered in Section~\ref{sec:tools}. 
\joe[drafts]{}
However, the quadtree is not built on the regions directly.  Rather, for
each region $R \in \cal R$, we store a \emph {representative point} $m$ that
lies somehow ``in the middle'' of $R$. 
We build search structures
over the quadtree which allow us to
quickly locate the quadtree cells containing relevant data. 
We answer point-location queries by locating the smallest quadtree cell containing
the query point and then searching the quadtree bottom-up for regions which
intersect this cell.
This approach allows us to handle input
described by arbitrary real numbers and to operate mostly on abstract
combinatorial objects.  We only require basic operations on our input: compare
two numbers, and find a bounding box around a small set of points (see
Appendix~\ref {sec:compmodel} for more details).

We first illustrate the main ideas of our approach in the simpler
one-dimensional version of the problem, in which we do not need any additional
search structure once we have located the correct quadtree cell.  In
Section~\ref{sec:1d}, we show how the dynamic balanced quadtree achieves worst-case
constant time local updates and logarithmic point location queries for intervals
in $\R^1$.
In Section~\ref{sec:2d} we solve the more complex two-dimensional problem,
in which we require more sophisticated search structures. 
We ``mark'' a small number of
carefully chosen quadtree cells near the representative point, and show how to adapt a
\emph{marked-ancestor} data structure to find the relevant regions once
we have located the correct quadtree cell.
By leveraging the
marked-ancestor tree and edge-oracle tree described in Section~\ref{sec:tools}, we are able to support queries in
$O(\log n)$ time and local updates in $O(\log \log n)$ time. We can also support
insertions and deletions as the composition of a query and local update. 


\ifFull
Our search structures require the assumptions made in Section~\ref{sec:prob}
when we defined \emph{local updates}. 
That is, we assume that the regions are fat and disjoint.
Realistic input models are intended 
for designing algorithms that are provably efficient in practice, and the fat-and-disjoint
model is ubiquitous 
    (see e.g. 
    \cite{bg-vrscdofo-06,ekns-ddsfoa-00,kos-ehsrosus-92} and citations therein).
%
Note that the fat-and-disjoint model is not a direct requirement of the quadtree, 
as the quadtree only
stores the representative points of regions. Rather, we leverage the model in order 
to bound the number of directions from which a region may overlap the query cell,
and thus facilitate fast queries in the marked-ancestor structure.

\fi


Thus we achieve our goal of maintaining a data structure with 
query time $Q(n)=\Theta(\log n)$ but local update time $U(n)=o(\log n)$. 
Note that for planar point location in rectilinear subdivisions 
$Q(n) = O(\log n)$ and $U(n)=O(\log n)$ can be achieved 
on a RAM by using the complex data structures of Blelloch~\cite{b-sedopl-08} or
Giora and Kaplan~\cite{gk-odvrs-09} and removing $R$  and re-inserting $R'$. 
However, we show that changing a region locally is more efficient than naively
removing a region and inserting a new one. 
%
Iacono and Langerman~\cite{il-dplfh-00} also give a solution which achieves $O(\log N)$ query
time and $O(1)$ update time if the regions are restricted to be disjoint axis-aligned fat hyper-rectangles with
coordinates drawn from a fixed universe $[N]$.
However, in our solution we are able to achieve sub-logarithmic local updates without requiring
that the regions be axis-aligned, rectangular, or limited precision. 
Moreover, our solution works on a real-valued pointer machine and does not require
hashing, bit-level manipulation, or even the floor operation (see
Appendix~\ref{sec:extensions}).


%
%



\section{Tools}
\label{sec:tools}

\ifFull
Before attacking the dynamic point location problem, we review several known and new concepts, techniques, data structures and notation that will help us.

\subsection {Preliminaries} 
\fi
\label {sec:prelim}

 \paragraph{Quadtrees.}
Let $B$ be an axis-aligned square.\footnote {We use the term \emph {square} to mean a $d$-dimensional hypercube, since our main focus is on $d=2$.}
    A \emph{quad\-tree} $T$ on $B$ is a hi\-erarchical
    decomposition of $B$ into smaller axis-aligned squares called quadtree
  \emph{cells}. 
  Each node $v$ of $T$ has an associated cell $C_v \subset \R^d$, and $v$ is
  either a leaf or has $2^d$ equal-sized children whose cells subdivide
  $C_v$~%
    \cite{deBergChvKrOv08,FinkelBe74,HarPeled11,Samet90}.  
\joe{Clarify definition of balance in terms of \emph{larger} neighbors and
replace share a boundary with share an edge or corner.}
  We denote the
  parent of a node $v$ by $\bar v$.
  A pair of cells are called \emph{neighbors} if they are interior disjoint and 
  meet at an edge or corner.  
    A leaf $v$ is  \emph{$\alpha$-balanced} if  
  $\alpha |C_v| \geq |C_u|$ for every larger neighbor $C_u$ of $C_v$.
  We say $T$ is \emph {$\alpha$-balanced} if every leaf in $T$ is
  $\alpha$-balanced.
    If $\alpha$ is a small constant (e.g., 2 or 4), 
    then we simply call the quadtree $T$ \emph{balanced}.
  
  Let $P \subset \R^d$ be a set of $n$ points contained in $B$. We say $T$ is a \emph {valid} quadtree for $P$ if every leaf of $T$ contains at most $1$ point of $P$. 
  We will be maintaining a valid quadtree for a certain set $P$, and require that the points and leaves that contain them are always connected by bidirectional pointers.
    It is known that quadtrees may have unbounded depth if $P$ has unbounded 
    spread,\footnote{The \emph{spread} of a point set $P$ is the ratio
      between the largest and the smallest distance between any two distinct
      points in $P$.}
    so in order to give any theoretical guarantees the concept is
    usually refined.
    Given a large constant $a$, an \emph {$a$-compressed} quadtree
    is a quadtree with additional \emph{compressed} nodes.
A compressed node $v$ has only one child $\tilde v$ with $|C_{\tilde v}|
  \leq |C_v|/a$ and such that $C_v \setminus C_{\tilde v}$ has no points from
  $P$.\footnote {
    Such nodes are also often called \emph {cluster}-nodes in the
  literature~%
    \cite{BernEpGi94,BernEpTe99,blmm-pipstse-10}.  
  }
  In the remainder, we
  assume for simplicity of exposition that $\tilde v$ is \emph {aligned} with $v$, that is, if we keep
  subdividing $C_v$ we will eventually create $C_{\tilde v}$.\footnote
  { While this assumption is realistic in practice, on a pure real-valued pointer machine
  it is not possible to align compressed nodes of arbitrary size difference in constant time.
  In Section~\ref {sec:ext-nofloor}, we show how to adapt the results to unaligned compressed nodes.
  }
        
  The compressed nodes of a quadtree $T$ cut the tree into a number of
  components that correspond to smaller regular (uncompressed) quadtrees. We
  say $T$ is $\alpha$-balanced if all these smaller trees are
  $\alpha$-balanced. It follows directly from
  Theorem 1 of Bern~\etal~\cite{BernEpGi94},
 that a balanced compressed quadtree of
  linear complexity exists for any set of points $P$.  
  \reviewer[2] {Can you provide a more exact reference for the balanced compressed
  quadtree (i.e., which Lemma/Theorem in [11] makes the claim)?}
\joe{(as an easy corollary of Theorem 1.) Is putting in the Theorem really
necessary?}
\maarten {I don't think so, actually. Usually people just refer to papers, not places within papers, unless it's a book or something like that. I would have kept it as it was, but I guess it's fine to change it (and who knows, we might get the same reviewer again and he might be happy about it).}

\paragraph {Static edge-oracle trees.}
Let $T$ be an abstract tree of size $|T|$ 
with constant maximum degree $d$.
Suppose that the nodes in the tree are given unique labels, and 
suppose that each edge $e \in T$ has an oracle which for any node label $x$ can answer
the following question:
``If we removed $e$ such that $T$ is split into two components,
which component would contain the node labeled $x$?''  
The edge-oracle tree is a search structure built over the edges of $T$ 
which allows us to navigate from
any node $u \in T$ to any other node $v \in T$ in 
$O(\log |T|)$ time 
and examines only 
$O(\log |T|)$ edges.
We can construct an edge-oracle tree for $T$ by recursively locating 
an edge which divides $T$ into two components of approximately equal size.  

The static version of this structure is similar to the well known
centroid-decomposition method for building a logarithmic height search structure
over an unbalanced tree. In fact, Arya \etal~\cite{amnsw-oaanns-94} used a
similar technique to support point location in a quadtree, but only
considered the static setting. 

\paragraph{Local updates.}
For a one-dimensional ordered list, data structures that can handle local (finger)
updates are well known. One of the simplest implementations on a pointer machine
is due to Fleischer~\cite{f-sbst-93}.


\paragraph{Marked-ancestor problem.}
Suppose we are given a simple path where some nodes in the path can be
marked, and we want to support the following query for any node $x$: 
``Which is the first marked node which comes after node $x$ in the path?''
and we also want to support updates where nodes can be marked or unmarked and
inserted into or deleted from the path. This is known as the \emph{marked
successor} problem. A natural generalization of this problem is to extend
support from paths to any rooted tree. Now the query we must support is 
``Which is the lowest marked ancestor of $x$ in the tree?''. This is known as
the \emph{marked-ancestor problem}. As in the marked successor problem, we also
want to support updates, in which nodes are marked or unmarked, and
insertions/deletions of nodes to/from the tree.
Alstrup~\etal{}~%
    \cite{ahr-map-98,ahr-map-98-conf}
gave the following
results for the marked-ancestor problem on a word-RAM.
\begin{lemma}
We can maintain a data structure over any rooted tree $T$ which supports
insertions and deletions of leaves in $O(1)$ amortized time, marking and
unmarking nodes in $O(\log \log n)$ worst-case time, and lowest marked
ancestor queries in $O(\log n / \log \log n)$ worst-case time.
\end{lemma}
\subsection{New Tools} \label {sec:imp}
We show how to maintain a \emph{dynamic} balanced compressed quadtree and 
 a \emph{dynamic} edge-oracle tree which supports local updates.
   \paragraph{Dynamic balanced quadtrees.} 
  A \emph {dynamic} quadtree is a data structure that maintains a quadtree $Q$
  on a point set $P$
  under insertion and deletion of points. In order to maintain a valid
  quadtree of linear size, we respond with \emph {split} and \emph {merge}
  operations respectively. A split operation takes a
  leaf $v$ of $Q$ and adds $2^d$ children to it; a merge operation takes $2^d$ leaves
  with a common parent and removes them.  
\ifFull
    Clearly, split and merge can be made to run in
  $O(1)$ time for quadtrees, since we are given the location in the quadtree where these operations are applied.
    
  In a dynamic compressed quadtree, we must consider the case where the node $v$ being split is a
  compressed node. In this case, $v$ gets $2^d$ new children, and $\tilde v$
  needs to be connected to the correct child. If the size factor
  is now less than $a$, this child gets further subdivided until the two
  components are merged.  
    A merge operation does the opposite.
    These operations can still be implemented in $O (1)$ time.
   
  In a balanced quadtree, after a split operation the balance may be
  disturbed, and we require additional cells to be split to restore the
  balance.  This operation may take $O (n)$ time in the worst case, if we want
  to maintain 2-balance, because the rebalancing operation can ``cascade''.
   However, if we only perform split operations, then we can maintain
   $4$-balance in $O(1)$ worst-case time per split.
\fi

  \begin{lemma}
  \label{lem:quadtree-balance}
  We can maintain $4$-balance
  in a dynamic compressed quadtree in $O(1)$ worst-case time per update.
  \end{lemma}
\ifFull
\reviewer[2]{The dynamic balancing of a quadtree is claimed in Lemma 2.2.
However, the proof of that lemma is much too vague for me. There is no place
where the authors clearly state what invariant is fulfilled by the quadtree.
The individual properties are introduced as they are needed, and it is not
clear to me how or why they can be maintained in each step. There is some
mention of a "regular pruning of the tree" without explanation of how
this is to be done. More worrisome, the proof does not seem to address 
the problem that is created by the interplay of compressed nodes with a 
balanced tree: when updating the tree, a compressed child may become part
of the regular tree. If this happens, I think that a lot of splitting may 
be necessary to reestablish the balancing condition. How is this this
situation handled?} 
\joe{by maintaining a ``buffer'' around a compresed node, as we discussed in
email. We need to mention this, but we only have space/time to sketch it.} 
 \fi
\begin{proof}(sketch)
We call a quadtree cell \emph{true} if its parent contains at least two
points of $P$ and it would therefore be present in any valid unbalanced quadtree, and
we call a quadtree cell a $B$-cell otherwise (i.e., it was only added to
maintain quadtree balance).
Figure~\ref {fig:qt-trueempty} shows an example.
\joe{we might need to more carefully state balance. Leaf cell with red point
right below purple cells is only 4-balanced with respect to purple cells. See
email discussion.}
\tweeplaatjes {qt-trueempty} {qt-insertion}
{ (a) A $2$-balanced quadtree on a set of points. True cells are shown in black,
$B$-cells in yellow.
  (b) An insertion could cause a linear ``cascade'' of cells needing to be split if we want to maintain $2$-balance. Therefore, we only split the direct neighbors (purple), which may now be only $4$-balanced. 
}
We will maintain the property that each true cell is $2$-balanced with respect
to its larger neighbors,
and every $B$-cell is $4$-balanced with respect to its larger neighbours.
\joe{neighboring \emph{true} cells, and 4-balanced wrt others?}

Let $C$ be a true quadtree cell which is $2$-balanced with respect to its neighbors.
When we split $C$, we examine the $3^d - 1$ neighbors of
$C$, and we split a larger neighbor $C'$ 
if the children of $C$ are not $2$-balanced with respect to $C'$.
 Thus we restore $2$-balance to $C$ at the cost of potentially inserting some
$B$-cells which are only $4$-balanced,
see Figure~\ref {fig:qt-insertion}.
\joe{What if the neighbor $C'$ was already a $B-cell$ and only 4-balanced wrt
its larger neighbors?}
 However, it takes two operations to
split a $B$-cell. First, we must insert a point into the $B$-cell, which
does not require a split since the cell was already split to maintain balance.
This changes the cell from a $B$-cell to a true cell. 
We also spend a constant amount of time examining each of the $O(1)$ neighbors
of the newly true cell, and splitting them if necessary so that the cell is now 
$2$-balanced with respect to its neighbors.  
\reviewer[2]{This is very vague. What do you mean by "enough time"? How exactly
  do you ensure the 2-balance (and maintain the 4-balance). I am not saying
  that what you claim is not correct, just that I find the explanation too
  hand-wavy.}

We may be splitting a
compressed node. Recall that if the size factor between a compressed node $v$
and it's child $\tilde v$ is less than $a$, then we continue to split $v$ a
constant number of times until the two components ``grow together''.
This case only requires a constant number of additional splits, and each split
can be handled in worst-case $O(1)$ time as before.
We maintain balance in the tree rooted at $\tilde v$ up to the level of $
\tilde v$, which ensures that no nodes more than a constant factor smaller than $v$ are on the outside, and only $O(1)$ work needs to be done to rebalance the tree.
\reviewer[2]{As mentioned above, I do not understand this. Don't you have
  to rebalance the tree when a compressed node and its child grow together?}
   \joe{Do we?}
   \maarten {Yes, but I don't think it should be a problem. The number of levels between the leaf of the big tree being split and the small tree inside it which gets merged with it, is constant. The small tree is already balanced, and doesn't need any further balancing. The big tree needs rebalancing because its leaf gets split, but since the outside cells of the small quadtree are of similar size to the leaf of the big quadtree, this is not really any different from a regular split.}


\joe{Trying out a version without lazy deletion.}
When we delete a $p$ from a cell $C$, we 
restore the quadtree to what it would be
had $p$ never been inserted, 
essentially ``undoing'' the insertion of $p$.
Since the original splitting and balancing only took $O(1)$ time, it clearly
only takes $O(1)$ to undo that splitting and balancing.
If $C$ was a $B$-cell, there is no
change. If $C$ was a true cell, and its parent $\bar C$ has smaller neighbors
which would become unbalanced if we merge $C$, 
then $\bar C$ may remain split and $C$ becomes a $B$-cell. Otherwise, 
we merge $C$. \joe[supposes]{If $\bar C$ has larger $B$-cell neighbors which are only split
because $\bar C$ was split, then these cells may also be merged. In total, we
only spend $O(1)$ time examining $\bar C$ its neighbors, and their neighbors. }
%
%

\joe{Why not?}
\maarten {I guess because we thought it's easier to just prune the tree regularly than to think about how to rebalance after deletions? I think it would probably work either way.}
\reviewer[2]{How and when is this "regular pruning" done? Is it spread out
  over the inserts/updates/queries? Is it a one-time operation? How is the
  work scheduled? Could you at least give a reference to some paper that uses
  a similar technique?}   
\end{proof}

\paragraph {Dynamic edge-oracle trees.}
%
There have been several recent results which 
generalize classic one-dimensional dynamic structures to a multidimensional setting by
combining classic techniques with a quadtree-style space decomposition. 
For example, the skip-quadtree~\cite{egs-skip-quad-08} combines
the quadtree and a skip-list, the quadtreap~\cite{mp-ddsars-10} 
combines a quadtree and a
treap, and the splay quadtree combines a quadtree with a splay tree~\cite{mp-sadsmps-12}.
However, surprisingly there are no multidimensional 
data structures which incorporate finger
searching techniques, i.e.
structures that are able to support both logarithmic queries and 
worst-case constant time local updates on a quadtree. 
\joe{New paper idea: ANN or approx range query with local updates.}
\ifFull
In the following we show how to build a dynamic edge-oracle 
tree which combines tree-decomposition and finger searching
techniques with a quadtree to support $O(\log n)$ queries and $O(1)$ local updates.
\else
We show how to build a dynamic edge-oracle 
tree which combines tree-decomposition and finger searching
techniques with a quadtree to support $O(\log n)$ queries and $O(1)$ local updates.
Due to space constraints, details are included in~\ref{app:eot}.
\fi
\ifFull
\begin{lemma} \label {lem:eot-height}
If $v$ is a leaf in an unweighted free tree $T$, 
then the edge incident to $v$ has height $O(1)$ in the
corresponding edge-oracle tree.
\end{lemma}
\begin{proof}
Recall that we construct the static edge-oracle tree for $T$ by recursively locating 
an edge which divides $T$ into two components of approximately equal size.  
Thus the edges are split in order to maintain
a balanced number of edges in each subtree of the edge-oracle tree. 
Since the edge adjacent to a leaf has 0 edges to one side of the split and at
least one edge on the other side of the split, these edges will not be chosen for
splitting by the algorithm until there are no other edge choices in the
sub-tree.
\end{proof}

\tweeplaatjes [width=.45\textwidth] {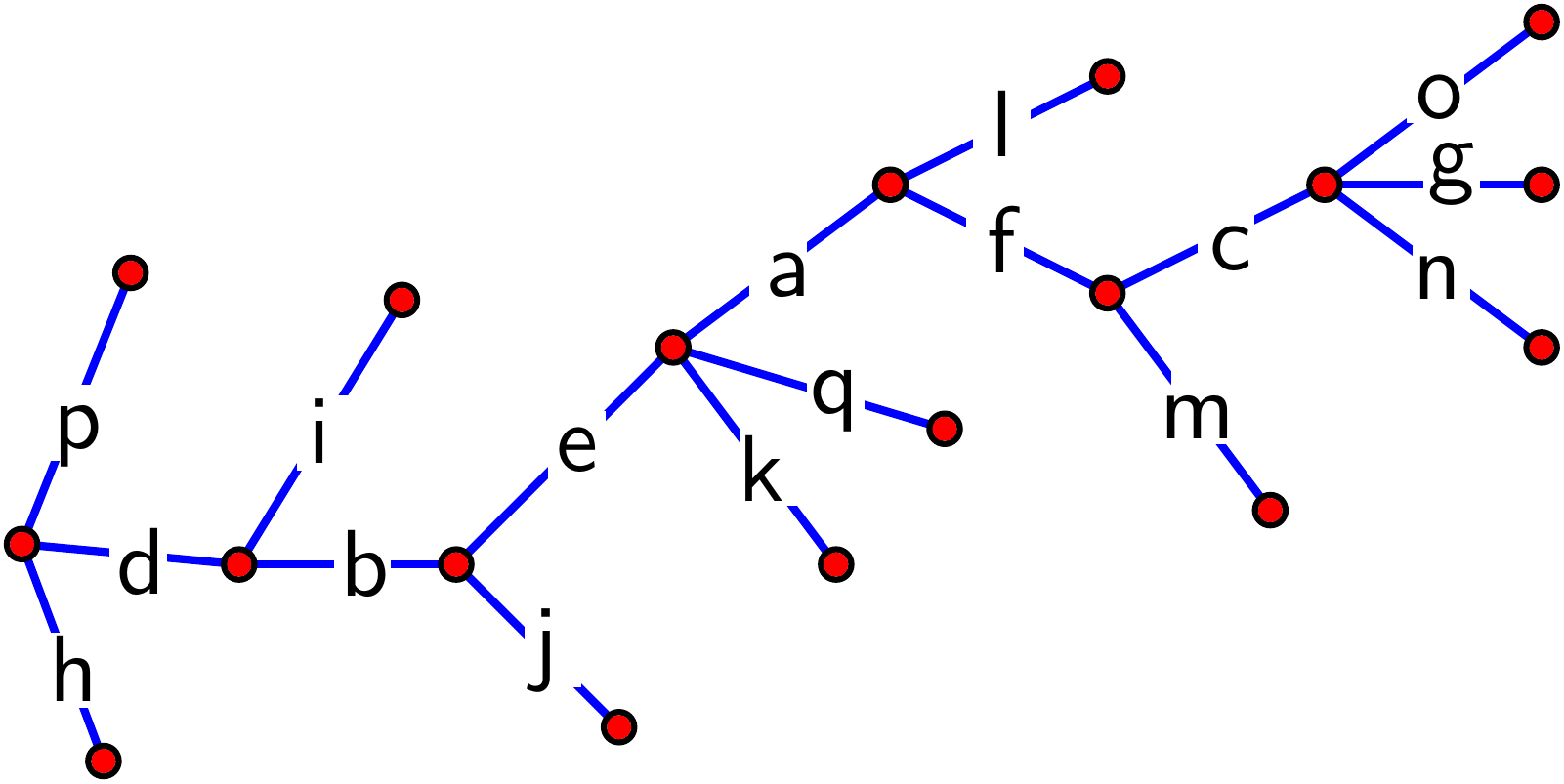} {edge-oracle-b}
{ A not necessarily balanced or rooted abstract tree $T$ (a) and its corresponding 
  edge-oracle tree (b). Labels on edges of $T$ match up with the label of the
  corresponding node in the edge-oracle tree. In the final structure we
  maintain the edge-oracle tree as a modified $(a,b)$-tree;
  small subtrees are maintained as buckets (linked-lists) to facilitate fast updates.
}

\begin{lemma} \label {lem:eot-dynamic}
Let $T$ be a tree subject to dynamic insertions and deletions of leaves. 
We can maintain an edge-oracle tree over $T$ in $O(1)$ worst case time per
local update. 
\end{lemma}
\begin{proof}
An insertion or deletion of a leaf and its associated edge in $T$ 
corresponds to an insertion or
deletion of a node in the edge-oracle tree. Since the location of the node is
known, and the height of the node is $O(1)$, we can borrow techniques 
from Fleischer~\cite{f-sbst-93} to perform updates in $O(1)$
time.
%
The techniques are surprisingly simple, and we only sketch them here.
We maintain the edge-oracle tree as an $(a,b)$-tree. However, we collapse each
subtree of size $\Theta(\log n)$ into a single pseudo-node called a \emph{bucket}. 
The original nodes
within the bucket are maintained in a simple linked-list. When
performing a query, we locate the correct bucket and iterate through the list for the correct 
original node in
$O(\log n)$ time. Given a pointer to an original node, an update is simply a $O(1)$
linked-list operation. If many nodes are inserted into the same bucket, then a
bucket may become too large. However, Fleischer shows how to distribute the rebuilding of
buckets over later updates, only spending $O(1)$ time per update, 
such that the size of each bucket never deviates significantly from
$\Theta(\log n)$. 
\end{proof}
%
\begin{lemma} \label {lem:eot-quadtree}
In a quadtree, an edge-oracle can be simulated in $O(1)$ time. 
\end{lemma}
\begin{proof}
In a quadtree, we are searching for the quadtree leaf which contains a
query point $q$. Each edge in a quadtree goes between a child cell and a parent
cell that contains it. If the child cell contains the query point, then the leaf
must be the child cell or one of its descendants, and the oracle returns
the corresponding component of the quadtree. Otherwise, the oracle returns the
other component of a quadtree. Since each quadtree cell is aware of its bounding
box, we can compare the query point with the child cell and return our answer in
constant time.
\end{proof}
\fi
\begin{lemma}\label{lem:edge-oracle}
Let $P$ be a set of $n$ points, and $Q$ be a balanced and compressed quadtree on $P$.
We can maintain $P$ and $Q$ in a data structure that supports $O(\log n)$ point location queries in $Q$, and local insertions and deletions of points in $P$ (i.e., when given the corresponding cells of $Q$) in $O(1)$ time. 
\end{lemma}
\ifFull
\begin{proof}
By Lemmas~\ref{lem:eot-dynamic} and~\ref{lem:eot-quadtree}, 
 we can maintain an edge-oracle tree over the compressed
quadtree which can find the unique quadtree cell containing a query point in $O(
\log n)$ time and respond to local updates in the quadtree in $O(1)$ time.
\end{proof}
\fi
\paragraph {Marked-ancestor trees.}
We show how to answer marked-ancestor queries on a
pointer-machine. Details are
given in Section~\ref{sec:MAT}.

\begin{lemma}
We can maintain a data structure over any rooted tree $T$ which supports
insertions and deletions of leaves in $O(1)$ amortized time, marking and
unmarking nodes in $O(\log \log n)$ worst-case time, and queries for the lowest marked
ancestor in $O(\log n)$ worst-case time.
All operations are supported on a pointer machine.
\end{lemma}


\section {One-Dimensional Case} \label {sec:1d}
\ifFull
To aid our exposition, we first present a solution to the one-dimensional version of the problem.
 Our data structure illustrates the key ideas of our approach while being
 significantly simpler than the two-dimensional version. 
\else
 Our 1D data structure illustrates the key ideas of our approach while being
 significantly simpler than the 2D version. 
\fi
 Note that in $\R^1$, our input set $\mathcal R$ of geometric regions is a set of
 non-overlapping intervals. 
 The difficulty of the problem comes from the fact that a local update may replace any interval by another interval of similar size at a distance related to that size; hence, it may ``jump'' over an arbitrary number of smaller intervals.
Our solution works on a pure Real-valued pointer machine, and achieves constant time updates.
  
  \subsection {Definition of the data structure}
Our data structure consists of two trees. The first is designed to facilitate
efficient updates and the second is designed to facilitate efficient queries. 
The update tree is a compressed quadtree on the center points of the intervals; 
The quadtree stores a pointer to each interval in the leaf that contains its center point.
We also augment the
tree with level-links, so that each cell has a pointer to its adjacent cells of
the same size (if they exist), and maintain balance in the quadtree as described
in Lemma~\ref{lem:quadtree-balance}. 
The leaves of the quadtree induce a linear size subdivision of the real line;
the query tree is a search tree over this subdivision\footnote
{Although we could technically also use a search tree directly on the original intervals, we prefer to see it as a tree over the leaves of the quadtree tree in preparation for the situation in $\R^2$.}
 that allows for fast point location
and constant time local updates.
We also maintain pointers between the leaves of the two trees, so that when we
perform a point location query in the query tree, we also get a pointer to
the corresponding cell in the quadtree, and given any leaf in the quadtree, we
have a pointer to the corresponding leaf in the query tree. 
Figure~\ref {fig:global} illustrates the data structure.

\eenplaatje [width=\textwidth]{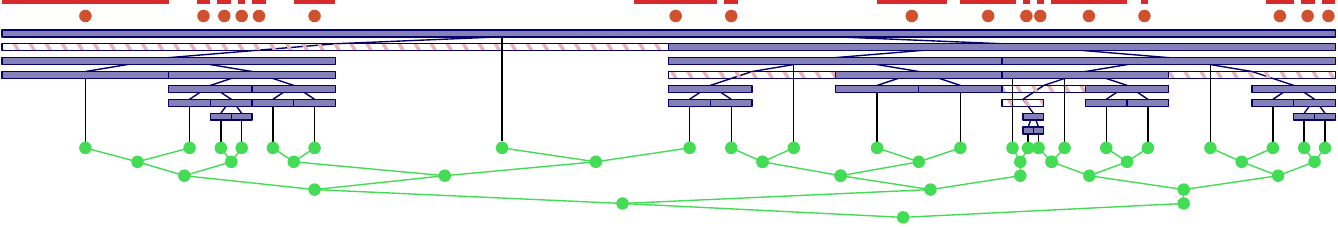} 
{ A set of disjoint intervals and their center points (red); a compressed
quadtree on the center points (blue); and a search tree on the leaves (or parts of internal cells not covered by children) of the quadtree (green).
}
\ifFull
\else
Details of the following results can be found in Appendix~\ref{app:1d}. 
\fi
\begin{lemma} \label{lem:1d-distance}
Let $I \in \mathcal R$ be an interval, and let $I'$ be another interval that is
$O(\rho)$-similar to $I$.
Suppose we are given a quadtree storing the midpoints of the intervals in $\mathcal R$ and a pointer to the leaf containing the midpoint of $I$.
Then we can find the leaf which contains the midpoint of $I'$ in $O(\log\rho)$ time.
\end{lemma}
\ifFull
\joe{In a pinch, for space constraints, I think we can only give the statement
of lemma and theorem. Give proof, handling updates and queries details in
appendix. Especially since we have most key ideas in the solution outline.}
\begin{proof}
Let $C$ be the quadtree leaf cell which contains the center point of $I$,
and let $C'$ be the quadtree cell which contains the center point of $I'$.
Observe that $I$ is at most four times as large as $C$: 
otherwise,
$I$ would completely cover the parent $\tilde C$ of $C$, but then no other intervals 
could have their center points in $\tilde C$ to cause $\tilde C$ to be split.
Similarly, $I'$ is at most four times as large as the new quadtree cell $C'$. 
Therefore, the distance between $I$ and $I'$ is proportional to the size of $C$ (and $C'$).
Since we
maintain balance in the quadtree according to Lemma~\ref{lem:quadtree-balance},
we can find $C'$ from $C$ by following $O(1)$ level-link and
parent-child pointers in the quadtree.
\end{proof}

\ifFull \subsection {Handling queries}
\else \paragraph{Queries.}
\fi
In a query, we are given a point $q$ and must return the interval in $\mathcal R$ 
that contains $q$.
We search in the query tree to find the quadtree leaf cell which contains
$q$ and its two neighboring cells in $O(\log n)$ time. Any interval $I$
which overlaps $q$ must have its center point in one of these three cells
(otherwise, there would be an empty cell between the cell containing $q$ and the cell containing the center point of $I$). We
compare $q$ with the intervals stored at these cells (if any) to
find the unique interval that contains $q$ or report that there is
no containing interval in $O(1)$ time. Thus the total time required by a query
is $O(\log n)$.   
  
\ifFull \subsection {Handling updates}
\else \paragraph{Updates.}
\fi
 
In an update, we are given a pointer to an interval $I \in \mathcal R$, and a new 
interval $I'$ that should replace $I$.
We follow pointers in the quadtree to find the new cell which contains the
center point. If $I$ and $I'$ are $O(1)$-similar, Lemma~\ref{lem:1d-distance} implies 
that the new cell is at most a constant number of cells away,
and we find the correct cell in $O(1)$ time. Then we remove the center point
from the old cell and insert it into the new cell, performing any compression or
decompression required in the quadtree. This only requires a constant number of
pointer changes in the quadtree and can be done in $O(1)$ worst-case time, and we
may also need to restore balance to the quadtree, which requires $O(1)$ worst-case
time by Lemma~\ref {lem:quadtree-balance}.  Finally, we follow
pointers from the quadtree to the query tree, and perform the corresponding
deletion and insertion in that tree, which by
Lemma~\ref{lem:edge-oracle} takes
only constant time. Thus, the entire update can be completed in
$O(1)$ worst-case time. 
  
Note that we can also insert or delete intervals from the data structure in
$O(\log n)$ time; we perform a query to locate where the interval belongs and a
local update to insert it or remove it. 
\fi
\begin{theorem}
\label{thm:1d-result}
We can maintain a linear size data structure over a set of $n$
non-overlapping intervals such that we can perform point location queries and
insertion and deletion of intervals in
$O(\log n)$ worst-case time and local updates in
$O(\log \rho)$ worst-case time.
  \end{theorem}
  \maarten {For consistency, we might also want to parameterise this result by $\rho$.}
  \joe{done}

\section {Two-Dimensional Case} \label {sec:2d}
\label{sec:fatdef}
We now focus our attention on disjoint fat regions in the plane. 
Intuitively, a fat region should not have any long skinny pieces.
We consider two types of fat regions which precisely capture this intuition:
\emph{thick} convex regions and \emph{wide} polygons. 
 We say $R$ is \emph {$\beta$-thick} if there exists a pair of concentric balls
 $I, O$
    with $I \subseteq R \subseteq O$ and $|O| \leq \beta |I|$, see Figure~\ref
    {fig:intro-betathick}.
Let $\delta \ge 1$.
A \emph {$\delta$-corridor} is a isosceles trapezoid whose slanted edges are at most $\delta$ times as long as its base.
A simple polygon $P$ is \emph{$\delta$-wide} if any isosceles trapezoid $T \subset P$ whose slanted
edges lie on the boundary of $P$ is a $\delta$-corridor~\cite {k-fpfcu-98},
see Figure~\ref {fig:intro-deltawide}.\footnote
{Many other notions of fatness exist in the literature. We chose to use thickness because it is basic and implied by most other definitions, and wideness because it will be convenient to use Theorem~\ref {thm:partition}.}
Note that any $\delta$-wide polygon $R$ of constant complexity is also $\beta$-thick, with $\beta \in \Theta(\delta)$.
%
%
\maarten {The dependence between wideness and thickness is a bit strange, in that it depends on the number of vertices of the wide polygon. Since we actually only use \emph {convex} regions in any of the lemmas (the non-convex ones will be chopped up before ever being inserted into the data structure), maybe we can state them in terms of convex thick regions. (I.e., I'd like them to be applicable to disks, which are not polygonal so not wide.)}
\tweeplaatjes {intro-betathick} {intro-deltawide}
{ (a) A $\beta$-thick region for $\beta = \frac {r_O} {r_I}$.
  (b) A $\delta$-wide region for $\delta = h/b$.
}
We will first solve the problem for convex thick regions, and then extend the result to 
non-convex wide polygons.
Analogously to the 1D case, we will store for each region $R \in \cal R$ a \emph {representative point} $p$ that lies somehow ``in the middle'' of $R$.
When the regions are $\beta$-thick, we will use the center point of the two concentric disks from the thickness definition as representative point.
  We denote the set of representative points of the regions in $\mathcal R$ by $P$.
  Let $T$ be the quadtree built over $P$.  
\joe{adding true vs B-cell distinction here, so we can use it in lemmas.}
We distinguish between \emph{true} cells, which are necessary in any valid
compressed quadtree over $P$, and $B$-cells, which may further subdivide a true cell
and are only added in order to maintain balance. We store each representative
point $m$ in $T$ according to the following rule:
Let $C_v$ be the smallest quadtree cell containing $m$. If $C_v$ is a true cell,
then $m$ is stored in $v$. If $C_v$ is a $B$-cell, then $m$ is stored in $u$, the lowest
(not necessarily proper) ancestor of $v$ in $T$ such that $|C_u| \geq |R| / (4\beta)$. 

%
Several new problems are introduced 
which were not
present in the 1D case. We briefly sketch how to address each of these
problems, and then present the complete solution.

\joe{swapped out a figure.} 
\tweeplaatjes [width=.35\textwidth]{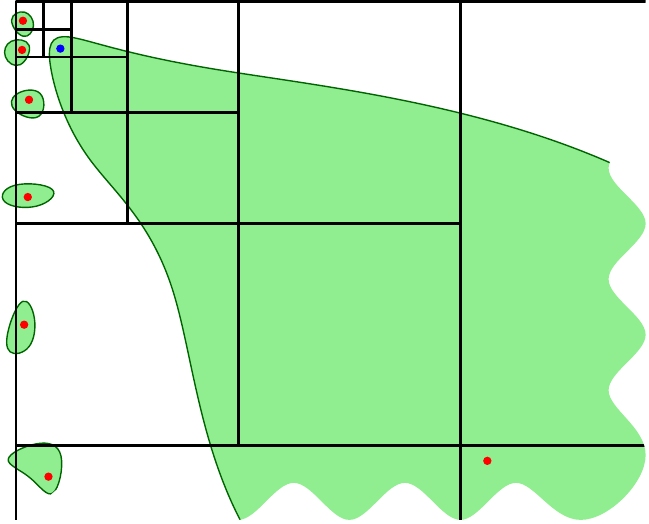} {circum-cover}
{ (a) The intersecting region could be stored a linear distance from the query cell
      (containing the blue point).
  (b) The number of regions which can intersect quadtree
leaf $C$ is at most $O(\beta)$, since each region blocks a $\Omega(1/\beta)$
fraction of a large
circle centered at $C$, by similar triangles.}


\paragraph{Linear distance.}
When performing a query in the one-dimensional case, the location in the quadtree
of any intersecting region is at most a
constant number of cells away. 
However, in the two-dimensional case, 
the location of an intersecting region may be up to a linear
number of cells away, as shown in Figure~\ref{fig:linearDistance}.
We solve this problem with some additional bookkeeping. Given a quadtree cell
$C_q$, we use two different strategies to locate regions intersecting $C_q$
depending on their size.
All regions of size at least $2\beta |C_q|$ will be located using 
a \emph{marked-ancestor} data structure:
an additional search structure which we explain in
more detail below. All regions of size less than $2 \beta |C_q|$ which intersect
$C_q$ will register a bidirectional pointer with $C_q$ using the following \emph{tagging}
strategy. 

Let $d$ be the smallest diameter of a quadtree cell such that $d \geq |R|/(4\beta)$.
Let $S_R$ be the set of quadtree cells $C$ which intersect $R$ and are either a leaf
or have size $|C| = d$.
%
All 
cells in $S_R$ will
be \emph{tagged} with a pointer to $R$.
Since the quadtree is balanced, given a pointer
to any cell in $S_R$, we can locate all cells in $S_R$ in $O(|S_R|)$ time. 
By the following lemma, $S_R$ must contain the 
cell containing the representative point of $R$.

 \begin{lemma}\label{lem:cell-size}
Let $R$ be a $\beta$-thick region stored by our data structure. 
If $C$ is the quadtree cell which stores the representative point of $R$, 
then $C$ has side length at least $\frac{|R|}{4\beta}$.
 \end{lemma}
\begin {proof}
If $C$ is a $B$-cell, then the claim is true by construction. Suppose $C$ is a
true cell.
  Let $m$ be the representative point of $R$.
  By the definition of thickness, there exists a disk $I \subseteq R$ centered at $m$ with $|I| \geq |R|/\beta$.
  $I$ contains no representative points of regions other than $R$.
  Let $C$ be the cell containing $m$. 
  Note that if $C$ contains $m$ and is significantly smaller
  than $|R|$, then $C$ must be completely
  contained in $I$. However, $C$ must be the largest quadtree cell completely
  contained in $I$, since if the parent $\bar C$ of $C$ in the quadtree is
  completely contained in $R$, then $\bar C$ would not have been further
  subdivided because $\bar C$ would contain no other points. 
  \joe{reviewer 1 complains: $\bar C$ may be split for rebalancing}
  \joe{Not anymore. Clarified true vs. B-cell}
  Therefore, $\bar C$ must have some portion outside of $I$ and must have size 
  larger than $|I|/2$. Thus the size of $C$ is at least 
  $|I|/4 \geq |R|/(4\beta)$.  
\end{proof}

Moreover, by the following lemma $|S_R| = O(\beta)$, 
and therefore, given the cell containing the
representative point of $R$ we can tag all cells in $S_R$ in
$O(\beta)$ time. 

\begin{lemma}\label{lem:2d-packing} 
Let $R$ be a $\beta$-thick region stored in
our data structure, and let $C$ be
quadtree cell that stores the representative point of $R$. Then there are at
most $O(\beta)$ quadtree cells of size $|C|$ required to cover $R$.  
\end{lemma}
\begin {proof} 
Let $I$ be the largest inscribed disk of $R$.
The boundary of $I$ touches the boundary of $R$ in two or three points. If two
points, then these are diametral on $I$, so $R$ is contained in a strip of width
$|I|$. If three points, then take the diametral points of these three points and
take the strips of width $|I|$ of these three pairs; $R$ is contained in the
union of these three strips. Now, if $R$ is beta-thick, the portion of the
strips it can be in is at most $\beta|I|$ long.
So, $R$ can be covered by $O(\beta)$ disks the size of $I$. Each such disk can
be covered by at most $O(1)$ cells of size $|C|$, by Lemma~\ref{lem:cell-size}.
Thus, $O(\beta)$ cells are required to cover $R$.
\end {proof}

\maarten {I'm still wondering whether this structure still makes sense, and if all of this shouldn't be somwhere under "definition of the data structure". (But I don't think we want to mess with it for SoCG.)}
\joe{I sure hope you're talking about the narrative structure of this section,
and not any of our data structures. If so, I think I'm tending to agree with
you. }
\maarten {Yes, that's the type of structure I am talking about. :)}

\paragraph{Linear overlap.} 
 In the one-dimensional case, we store only the center points of our regions,
 and the number of regions that overlap any quadtree cell is at most three.  In
 two dimensions, it appears that we may have a large number of small regions
 that intersect a quadtree cell.  However, we show in the following lemma
 that this is not the case. 

\begin{lemma}\label{lem:bounded-overlap}
The number of 
$\beta$-thick convex regions intersecting any balanced quadtree leaf is 
$O(\beta)$. 
\end{lemma}
\begin {proof}
Let $R_C$ be the set of thick convex regions that intersect the boundary of 
leaf $C$, and 
let $r$ be the radius of a large disk $D$ containing all regions in $R_C$.
For each region $R_j \in R_C$ 
there exists a disk $I_j \subseteq R_j$ with center $m_j$ such that $|I_j| \geq |R_j| / \beta$.
Moreover, since each region $R_j$ is convex, 
it must contain a triangle 
consisting of the diameter of $I_j$ 
and some point $p_j \in R_j \cap C$. Each of the four sides of $C$ can ``see'' at most $\pi r$ of the perimeter
of $D$. However,  by a similar
triangles argument
each triangle 
must block the line of sight from one or
more sides to at least $\Theta(r / \beta)$ of the
perimeter (see Figure~\ref{fig:circum-cover}).
Thus, since the regions are convex and disjoint, 
the number of regions in $R_C$ is at most $O(\beta)$. 
\end {proof}

  \subsection {Definition of the data structure} 
    At the core, our data structure is similar to the one-dimensional data structure
    described above: we have a spacial tree, which allows for efficient updates,
    and a search tree, which allows for efficient searching over the quadtree.
    However, our data structure is augmented to address the problems introduced by the
    two-dimensional case.
    We maintain a dynamic balanced quadtree $Q$ over $P$, which we
    augment to support \emph {mark} and \emph {unmark} operations and marked-ancestor
    queries,
    and we maintain a dynamic edge-oracle tree 
    on the edges of $Q$.
 \joe{Reviewer 4 wants sentence summarizing structure with parameters used,
 number of MAT, and how much space they use.} 
\paragraph {Marked-ancestor tree.}

Suppose we are given an angle $\phi$ which divides $2\pi$ (i.e., $k\phi = 2\pi$), and consider the set of angular intervals $\Phi_i = [i \phi, (i+1) \phi]$ (modulo $2\pi$), for integers $1 \leq i \leq k$.
For each quadtree cell $C$ of $Q$ with center point $c$, we define the wedge $W_{C}^i$
centered at $c$ and with opening angle $\phi$ to be the union of all halflines from $c$ in a direction in $\Phi_i$. 
Let ${\cal W}_C = \{W_{C}^i \mid 1 \le i \le k\}$; note that ${\cal W}_C$ partitions $\R^2$ into $k$ wedges.

For each $1 \le i \le k$, let $T_i$ be a marked-ancestor structure on $Q$.
We mark a cell $C$ in $T_i$ if and only if there is a region $R \in \cal R$ of size
$2\beta|C| \le |R| < 4\beta|C|$ that intersects $C$, and such that the center
point of $R$ lies in $W_{C}^i$.

When doing a query, we will only look at the first marked ancestor in each $T_i$.
Lemma~\ref{lem:block} captures the essential property of the regions which
enables this strategy. First, we need the following claim.

\begin {claim} \label {cla:lines}
  Let $\beta$ be given and set $\phi = \frac{2\pi}{\lceil13\beta\rceil}$. 
 Let $C$ be a cell that is marked in $T_i$ by a $\beta$-thick region $R$.
  Let $L_{C}^i$ be the set of lines that start in $C$, and have a direction in $\Phi_i$.
  Then every line in $L_{C}^i$ intersects $R$.
\end {claim}
\tweeplaatjes {fat-wedge} {lem6-proof} {(a) Illustration of Claim~\ref {cla:lines}. (b) Illustration of Lemma~\ref {lem:block}.}
\begin {proof}
  Let $m$ be the representative point of $R$.
  Since $R$ is $\beta$-thick, there exist disks $I \subseteq R \subseteq O$
  centered at $m$ with $|O|/|I| \le \beta$. 
  Since $R$ caused $C$ to be marked, $O$, must intersect $C$, 
  and $m$ must lie in $W_{C}^i$. See Figure~\ref {fig:fat-wedge}.
  
  Now, we need that $I$ intersects all lines in $L_{C}^i$.
  The distance from $m$ to $C$ is at most $\frac12|O| \le \frac\beta2|I|$.
  Then, the distance from $m$ to the far edge of $W_{C}^i$ is at most 
  $\frac\beta2|I|\sin\phi$,
  and the distance to the far edge of $L_{C}^i$ is at most $\frac\beta2|I|\sin\phi + \frac12|C|$.
  Since $|R| \ge 2\beta|C|$, we know that $|C| \le \frac12|I|$.
  Using $\phi = \frac{2\pi}{13\beta}$ implies $\beta \sin\phi \le
  \frac{2\pi}{13} < \frac{1}{2}$.
  Combining these, we see that
  $|I| \ge \beta|I|\sin\phi + |C|$,
  so, $I$ blocks all lines in $L_{C}^i$.
\end {proof}

\begin {lemma} \label {lem:block}
Let $C_1$ be a cell that is marked in $T_i$ by a convex and $\beta$-thick region $R_1$, and let $C_2$ be a descendant of $C_1$ that is marked in $T_i$ by a convex and $\beta$-thick region $R_2$.
Then there cannot be a descendant $C_3$ of $C_2$ that intersects $R_1$.
\end {lemma}
\begin {proof}
Let $R_2$ and $R_1$ be convex fat regions which mark cells $C_2$ and $C_1$
respectively. Then there is a point $p_2 \in R_2 \cap C_2$. 
Suppose for contradiction that $R_1$ intersects $C_3$; that is, there exists a
point $p_1 \in R_1 \cap C_3$.  Let $r$ and $s$ be two parallel rays from $p_1$
and $p_2$ in 
some direction $\phi \in \Phi_i$.
Note that rays $r$ and $s$ are
both in $L_{C_2}^i$.  Therefore each ray must intersect both $R_1$ and $R_2$ by
Claim~\ref{cla:lines}.  Since each region $R_1$ and $R_2$ is convex, their
intersection with each ray $r$ (or $s$) is a single line segment, denoted $r_1$
and $r_2$ ($s_1$ and $s_2$) respectively.  Moreover, since $R_1$ and $R_2$ are
disjoint, the segments $r_1$ and $r_2$ ($s_1$ and $s_2$) are also disjoint (see
Figure~\ref{fig:lem6-proof}). 

Since $p_1 \in R_1$, $r_1$ must come before $r_2$ on the ray $r$.  Similarly,
$s_2$ must come before $s_1$ on the ray $s$.  Moreover, $R_1$ is convex, and
thus the convex quadrilateral defined by $r_1, s_1$ is completely contained in
$R_1$, and likewise $r_2, s_2 \subseteq R_2$. These two quadrilaterals must intersect,
which is a contradiction because $R_1$ and $R_2$ are disjoint.  Therefore there
is no point $p_1 \in R_1 \cap C_3$.
\end {proof}


\ifFull \subsection {Handling queries}
\else \paragraph{Queries.}
\fi
Given a query point $q$, we want to find out which region (if any) contains $q$.
We begin by performing a point location query for $q$ in the quadtree $Q$. By
Lemma~\ref{lem:edge-oracle} we can find the leaf cell $C$ in the
quadtree which contains $q$ in $O(\log n)$ time using the edge-oracle tree.

By Lemma~\ref{lem:bounded-overlap}, there can only be $O(\beta)$
regions which intersect $C$.
All regions of size at most $2\beta|C|$ will have tagged $C$ with a pointer to
themselves, and are immediately available from $C$.
Moreover, we can find all regions of size at least $2\beta |C|$ in $O (\beta \log n)$ time 
by querying the marked-ancestor structures.
We compare each region to our query point, and determine which region (if any)
intersects the query point in $O(\beta)$ time. Thus, we can answer the query
in total time $O(\beta \log n)$.

\ifFull \subsection {Handling updates}
\else \paragraph{Updates.}
\fi
 We only store the representative points of the regions in the quadtree. Thus, when performing a
 local update, it is sufficient to find the new location for the region's representative point,
 and then update the quadtree, tags, marked-ancestor trees, and 
 edge-oracle trees accordingly.

\ifFull \paragraph {Locating the new representative point.} \fi
Given a pointer to a region $R$,
we replace it by another region $R'$ that is $\rho$-similar to $R$
for any arbitrary parameter $\rho \ge 1$.
Let $p$ and $p'$ be the representative points of $R$ and $R'$, respectively. 
We find the leaf cell of $Q$ containing $p'$ by going up in the quadtreee until the size of the cell we are in is similar to the distance to $p'$, then using level-links to find the ancestor of $p'$ of similar size, and then going back down.

\begin {lemma}\label{lem:param-dist}
The distance in $Q$ between the leaf $C$ containing $p$ and the leaf $C'$ containing $p'$
is at most $O(\log (\rho\beta))$.
\end {lemma}
\begin {proof}
Recall that by definition, 
$|R \cup R'| \leq \rho \min\{|R|, |R'|\}$, and by Lemma~\ref{lem:cell-size}, each region is stored in a quadtree cell
proportional to its size, i.e. $|C| \geq \frac{|R|}{4\beta}$. 
Thus, $|C| \geq \frac{|R \cup R'|}{4\beta \rho}$, and likewise for $|C|'$. 
\joe{reviewer 1 caught a nice typo here, but it made it look like the proof was
incorrect. fixed.}
Hence, to find $C'$ from $C$, we
move up at most $\log(\beta \rho)$ levels in the quadtree to find a cell
of size $\Omega(|R \cup R'|)$, then follow $O(1)$ level-link pointers to find a large cell
containing $p'$. Finally, we move down at most $\log(\beta \rho)$ levels to find
$C'$.
\end {proof}

\ifFull \paragraph {Updating the quadtree.} \fi
We must also update the quadtree to reflect the new position
of the representative point. 
By Lemma~\ref{lem:quadtree-balance},
we can delete $p$, insert $p'$, and
perform the corresponding 
rebalancing of the quadtree in $O(1)$ worst case time.

\ifFull \paragraph {Updating the auxiliary structures.} \fi
A local update replaces an old region $R$ by a new region $R'$ which is $\rho$-similar to $R$,
but may overlap different quadtree cells than $R$.
Therefore we may require updates to the marked-ancestor structure.
Let $C$ be the quadtree cell containing $R$'s representative point.
After the update, $R'$ must only intersect $O(\beta)$ 
quadtree cells which are similar in size to $C$ by Lemma~\ref{lem:2d-packing}. 
For each of these
cells, we test the direction of the representative point of $R'$ and mark it in the
corresponding marked-ancestor tree. We
also unmark cells which corresponded to the old region $R$. 
These updates can be performed in $O(\log \log n)$ time per marked-ancestor structure.
We must also remove tags from all cells in $S_R$ and add tags to cells in
$S_{R'}$. However, given $C$ and $C'$, this takes $O(\beta)$
time by Lemma~\ref{lem:2d-packing}. 
By Lemma~\ref{lem:edge-oracle} we can also update the edge-oracle tree in
$O(1)$ time.
\maarten {Say something about the updating the tagged cells here?}
\joe{I described that in the previous paragraph, under updating the quadtree,
but it could just as easily go here. Actually, that might make more sense. Moved
it.}

\begin{theorem}
 \label{thm:2d-result}
  A set of $n$ disjoint convex $\beta$-thick objects of
  constant combinatorial complexity in $\R^2$ can be maintained
  in a $O(\beta n)$ size data structure that supports 
  insertion, deletion and
  point location queries in $O(\beta \log n)$ time,
  and $\rho$-similar updates in $O(\beta \log \log n + \log (\beta\rho))$ time.
  All time bounds are worst-case, and the data structure can be implemented on a real-valued
  pointer machine.
\end{theorem}

\ifFull
\subsection {Non-convex regions}
\fi

We can extend the result to non-convex fat regions, by cutting them into convex pieces. This approach only works for polygonal objects, since non-polygonal objects cannot always be partitioned into a finite number of convex pieces. For polygonal objects, we use a theorem by van Kreveld:

\begin {theorem} [from \cite {k-fpfcu-98}] \label {thm:partition}
  A $\delta$-wide simple polygon $P$ with $n$ vertices can be partitioned in $O (n \log^2 n)$ time
  into $O (n)$ $\beta$-wide quadrilaterals and triangles, where $\beta =
  \min\{\delta,1-\frac12\sqrt3\}$.
\end {theorem}

We conclude:
\begin{theorem}
  A set of $n$ disjoint polygonal $\delta$-wide objects of
  constant combinatorial complexity in $\R^2$ can be maintained
  in a $O(\delta n)$ size data structure that supports 
  insertion, deletion and
  point location queries in $O(\delta \log n)$ time,
  and $\rho$-similar updates in $O(\delta \log \log n + \log (\delta\rho))$ time.
  All time bounds are worst-case, and the data structure can be implemented on a real-valued
  pointer machine.
\end{theorem}

\ifFull
Note that $\alpha,\beta$-covered objects are $O(\min\{\alpha,\beta\})$-thick and polygonal $\alpha,\beta$-covered objects are $O(\min\{\alpha,\beta\})$-wide, so our results apply to such objects as well.
\fi

\section {Discussion} \label {sec:discussion}
We have shown that we can maintain a set of intervals in $\R^1$ or
disjoint fat regions $\R^2$ 
in a data structure that supports $O(\log n)$ point location queries,
and local updates in $\R^1$ in $O(1)$ time and
in $\R^2$ in $O(\log \log n)$ time respectively.
These results are the first of their kind in a geometric setting.
\joe{cutting some content}
Still, several gaps remain, and there are many open problems left for future research. 
 
\ifFull
We show that the fatness restriction is necessary given our current definition of locality.
However, for non-fat objects, the definition seems to be too powerful:
if all regions are skinny but homothetic, for example, we could solve the problem simply by scaling the plane in one direction. As soon as the regions have different orientations, however, this simple solution no longer works.
It would be interesting to investigate alternative, more restrictive definitions of similarity that capture this effect, and analyze to what extend local updates on non-fat objects can then be supported.

Also, it is unclear whether the disjointness condition is necessary. 
While the restriction is very natural in applications where the regions represent physical objects, it would be useful to be able to handle some restricted amount of 
\maarten {Currently, we don't define or use the word ``ply'' at all in the main body.} 
overlap when the regions represent imprecision.
However, it appears to be hard to extend our approach in this setting: even simply keeping a constant number of copies of our data structure does not work, because now one needs to assign regions to layers on the fly, which appears to be non-obvious.
\joe{Maybe we should include a result from one of the other fatness papers. The
reason they need fatness is to bound the complexity of the union of the objects.
But even for fat regions, if they overlap, the union is super-linear. I don't
know if that matters enough to mention it.}

\joe{We might want to cut previous paragraphs in this section almost entirely.
They reflect an older version of this paper. I don't know that they add much.}
\joe{Should observe the historical trend? that simplifying
fatness assumption is applied at first in order to tackle problem for the first
time, and subsequent research is able to relax assumptions to some degree. (e.g.
with the union complexity of fat regions line of research. First they have very
tight definition of fatness, then broader definition that includes more regions,
and finally only require that regions be not-too-dense}

Finally, perhaps 
\else
Perhaps
\fi
the most intriguing question left open regards the update complexity itself. 
While the $O(\log \log n)$ update time in the $2$-dimensional case is
sublogarithmic, it is not clear whether this is the right bound, or whether
constant time updates might be possible, as in the $1$-dimensional case.

\joe{We should consider including some of the content from unbounded skinniness paragraph
in the appendix here. e.g. it is not clear how to define similarity, it would be
interesting to investigate..., etc.}
\joe{I'm 90\% sure we can handle $R^3$ with very little adaptation. Should we still leave
this as an open question?}
\maarten {I agree, it probably would just work. Maybe we should be claiming it. Let's just not mention 3D then.}

\joe{We used to clearpage before acknowledgements, but I'd like to try to fit it
in.}
\small
\section*{Acknowledgments}
\joe{Cut this section out for space?}
\maarten {My grant number needs to be in the proceedings version if it gets accepted, but for submitting we could cut it.}
Work on this paper has been partially supported by the Office of
Naval Research under MURI grant N00014-08-1-1015.
M.L. is further supported by the Netherlands Organisation for Scientific Research (NWO) under grant 639.021.123.

\clearpage
\bibliographystyle {abbrv}
\bibliography {dapijd}
\normalsize

%

\newpage
\appendix

\section {Model of Computation} \label {sec:compmodel}

We wish to store regions described by arbitrary real numbers in our data structure.
In computational geometry, the standard model of computation is the real RAM model.
A real RAM is a random access machine with additional support for real number arithmetic.
In particular, one works with an abstract machine with an array of memory cells, each of which can either store a single real number, or an integer.
One is allowed to perform basic algebraic operations on real numbers in constant time, and to do integer arithmetic and use integers as address pointers as on a standard random access machine.
Additionally, one sometimes allows conversion from real numbers to integers (e.g., using a \emph {floor} operation): this is justified by the fact that in practice, real numbers are approximated by floating-point numbers on which the floor operation is trivial to execute, but controversial because it breaks the internal consistency of the computation model.
Similar to the real RAM, we may consider a real-valued pointer machine, which is a pointer machine with additional support for real number arithmetic. Like the real RAM, it has memory cells which store real numbers or integers; however, here the integers cannot be manipulated at all, they only function as abstract ``pointers'' to other memory cells.

In our data structure and the associated algorithms, we need to be able to compare real numbers to integers.
Furthermore, to build a quadtree, we need an operation that, given a set of real numbers, provides us with an interval that contains all numbers in the set, and whose length is approximately the difference between the largest and smallest numbers in the set. 
On limited-precision machines supplied with a floor operation, we can easily find the smallest interval containing the numbers whose length and end points are powers of $2$,
and use this to keep the quadtree aligned with the number system of the machine.
In the description of our results, we assume that this is the case.
However, if we are not able to convert real numbers to integers, as we would not be on a pure real RAM or real-valued pointer machine, we can also simply return the interval spanned by the smallest and largest element of such a set, and use real arithmetic to subdivide the interval and construct a quadtree.
For this, we additionally need to be able to compare real numbers to each other, to add and subtract them, and to divide them by $2$.
In Appendix~\ref {sec:ext-nofloor} we describe how to deal with compressed quadtrees on a pure real-valued pointer machine, in which no floor operation is available.
All other machinery operates on the combinatorial tree. We do need to manipulate integers (i.e., pointers) in order to use the marked-ancestor data structure by Alstrup~\etal{}~\cite{ahr-map-98}.
In Appendix~\ref {sec:MAT} we describe how to adapt this structure to a pointer machine, at the cost of an increase in query time (but since our queries are dominated by point location anyway, this does not affect our final result).

\ifFull
\else
\section{Edge-Oracle Trees}
\label{app:eot}
%
There have been several recent results which 
generalize classic one-dimensional dynamic structures to a multidimensional setting by
combining classic techniques with a quadtree-style space decomposition. 
For example, the skip-quadtree~\cite{egs-skip-quad-08} combines
the quadtree and a skip-list, the quadtreap~\cite{mp-ddsars-10} 
combines a quadtree and a
treap, and the splay quadtree combines a quadtree with a splay tree~\cite{mp-sadsmps-12}.
However, surprisingly there are no multidimensional 
data structures which incorporate finger
searching techniques, i.e.
structures that are able to support both logarithmic queries and 
worst-case constant time local updates on a quadtree. 
In the following we show how to build a dynamic edge-oracle 
tree which combines tree-decomposition and finger searching
techniques with a quadtree to support $O(\log n)$ queries and $O(1)$ local updates.

\begin{lemma} \label {lem:eot-height}
If $v$ is a leaf in an unweighted free tree $T$, 
then the edge incident to $v$ has height $O(1)$ in the
corresponding edge-oracle tree.
\end{lemma}
\begin{proof}
Recall that we construct the static edge-oracle tree for $T$ by recursively locating 
an edge which divides $T$ into two components of approximately equal size.  
Thus the edges are split in order to maintain
a balanced number of edges in each subtree of the edge-oracle tree. 
Since the edge adjacent to a leaf has 0 edges to one side of the split and at
least one edge on the other side of the split, these edges will not be chosen for
splitting by the algorithm until there are no other edge choices in the
sub-tree.
\end{proof}

\tweeplaatjes [width=.45\textwidth] {edge-oracle-a} {edge-oracle-b}
{ A not necessarily balanced or rooted abstract tree $T$ (a) and its corresponding 
  edge-oracle tree (b). Labels on edges of $T$ match up with the label of the
  corresponding node in the edge-oracle tree. In the final structure we
  maintain the edge-oracle tree as a modified $(a,b)$-tree;
  small subtrees are maintained as buckets (linked-lists) to facilitate fast updates.
}

\begin{lemma} \label {lem:eot-dynamic}
Let $T$ be a tree subject to dynamic insertions and deletions of leaves. 
We can maintain an edge-oracle tree over $T$ in $O(1)$ worst case time per
local update. 
\end{lemma}
\begin{proof}
An insertion or deletion of a leaf and its associated edge in $T$ 
corresponds to an insertion or
deletion of a node in the edge-oracle tree. Since the location of the node is
known, and the height of the node is $O(1)$, we can borrow techniques 
from Fleischer~\cite{f-sbst-93} to perform updates in $O(1)$
time.
%
The techniques are surprisingly simple, and we only sketch them here.
We maintain the edge-oracle tree as an $(a,b)$-tree. However, we collapse each
subtree of size $\Theta(\log n)$ into a single pseudo-node called a \emph{bucket}. 
The original nodes
within the bucket are maintained in a simple linked-list. When
performing a query, we locate the correct bucket and iterate through the list for the correct 
original node in
$O(\log n)$ time. Given a pointer to an original node, an update is simply a $O(1)$
linked-list operation. If many nodes are inserted into the same bucket, then a
bucket may become too large. However, Fleischer shows how to distribute the rebuilding of
buckets over later updates, only spending $O(1)$ time per update, 
such that the size of each bucket never deviates significantly from
$\Theta(\log n)$. 
\end{proof}
%
\begin{lemma} \label {lem:eot-quadtree}
In a quadtree, an edge-oracle can be simulated in $O(1)$ time. 
\end{lemma}
\begin{proof}
In a quadtree, we are searching for the quadtree leaf which contains a
query point $q$. Each edge in a quadtree goes between a child cell and a parent
cell that contains it. If the child cell contains the query point, then the leaf
must be the child cell or one of its descendants, and the oracle returns
the corresponding component of the quadtree. Otherwise, the oracle returns the
other component of a quadtree. Since each quadtree cell is aware of its bounding
box, we can compare the query point with the child cell and return our answer in
constant time.
\end{proof}
\begin{lemma}\label{lem:edge-oracle-app}
Let $P$ be a set of $n$ points, and $Q$ be a balanced and compressed quadtree on $P$.
We can maintain $P$ and $Q$ in a data structure that supports $O(\log n)$ point location queries in $Q$, and local insertions and deletions of points in $P$ (i.e., when given the corresponding cells of $Q$) in $O(1)$ time. 
\end{lemma}
\begin{proof}
By Lemmas~\ref{lem:eot-dynamic} and~\ref{lem:eot-quadtree}, 
 we can maintain an edge-oracle tree over the compressed
quadtree which can find the unique quadtree cell containing a query point in $O(
\log n)$ time and respond to local updates in the quadtree in $O(1)$ time.
\end{proof}
\fi

\ifFull
\else
\section{Details of the One-Dimensional Case}
 \label {app:1d}
To aid our exposition, we first present a solution to the one-dimensional version of the problem.
 Our data structure illustrates the key ideas of our approach while being
 significantly simpler than the two-dimensional version. 
 Note that in $\R^1$, our input set $\mathcal R$ of geometric regions is a set of
 non-overlapping intervals. 
 The difficulty of the problem comes from the fact that a local update may replace any interval by another interval of similar size at a distance related to that size; hence, it may ``jump'' over an arbitrary number of smaller intervals.
Our solution works on a pure Real-valued pointer machine, and achieves constant time updates.
  
  \subsection {Definition of the data structure}
Our data structure consists of two trees. The first is designed to facilitate
efficient updates and the second is designed to facilitate efficient queries. 
The update tree is a compressed quadtree on the center points of the intervals; 
the quadtree stores a pointer to each interval in the leaf that contains its center point.
We also augment the
tree with level-links, so that each cell has a pointer to its adjacent cells of
the same size (if they exist), and maintain balance in the quadtree as described
in Lemma~\ref{lem:quadtree-balance}. 
The leaves of the quadtree induce a linear size subdivision of the real line;
the query tree is a search tree over this subdivision\footnote
{Although we could technically also use a search tree directly on the original intervals, we prefer to see it as a tree over the leaves of the quadtree tree in preparation for the situation in $\R^2$.}
 that allows for fast point location
and constant time local updates.
We also maintain pointers between the leaves of the two trees, so that when we
perform a point location query in the query tree, we also get a pointer to
the corresponding cell in the quadtree, and given any leaf in the quadtree, we
have a pointer to the corresponding leaf in the query tree. 
Figure~\ref {fig:global} illustrates the data structure.

\ifFull 
\eenplaatje {global} 
{ A set of disjoint intervals and their center points (red); a compressed
quadtree on the center points (blue); and a search tree on the leaves (or parts of internal cells not covered by children) of the quadtree (green).
}
\fi
\begin{lemma} \label{lem:app-1d-distance}
Let $I \in \mathcal R$ be an interval, and let $I'$ be another interval that is
$O(\rho)$-similar to $I$.
Suppose we are given a quadtree storing the midpoints of the intervals in $\mathcal R$ and a pointer to the leaf containing the midpoint of $I$.
Then we can find the leaf which contains the midpoint of $I'$ in $O(\log\rho)$ time.
\end{lemma}
\begin{proof}
Let $C$ be the quadtree leaf cell which contains the center point of $I$,
and let $C'$ be the quadtree cell which contains the center point of $I'$.
Observe that $I$ is at most four times as large as $C$: 
otherwise,
$I$ would completely cover the parent $\tilde C$ of $C$, but then no other intervals 
could have their center points in $\tilde C$ to cause $\tilde C$ to be split.
Similarly, $I'$ is at most four times as large as the new quadtree cell $C'$.
Recall that by definition $|I \cup I'| \leq \rho \min\{|I|, |I'|\}$. Thus, to
find $C'$ from $C$, we move up at most $O(\log \rho)$ levels in the quadtree to
find a cell of size $\Omega(|I \cup I'|)$, then follow $O(1)$ level-link
pointers to find a large cell containing the center point of $I'$, then move
down at most $O(\log \rho)$ levels to find $C'$.
\end{proof}

\subsection {Handling queries}
In a query, we are given a point $q$ and must return the interval in $\mathcal R$ 
that contains $q$.
We search in the query tree to find the quadtree leaf cell which contains
$q$ and its two neighboring cells in $O(\log n)$ time. Any interval $I$
which overlaps $q$ must have its center point in one of these three cells
(otherwise, there would be an empty cell between the cell containing $q$ and the cell containing the center point of $I$). We
compare $q$ with the intervals stored at these cells (if any) to
find the unique interval that contains $q$ or report that there is
no containing interval in $O(1)$ time. Thus the total time required by a query
is $O(\log n)$.   
  
\subsection {Handling updates}
 
In an update, we are given a pointer to an interval $I \in \mathcal R$, and a new 
interval $I'$ that should replace $I$.
We follow pointers in the quadtree to find the new cell which contains the
center point. If $I$ and $I'$ are $\rho$-similar, Lemma~\ref{lem:app-1d-distance} implies 
and we find the correct new cell in $O(\log \rho)$ time. Then we remove the center point
from the old cell and insert it into the new cell, performing any compression or
decompression required in the quadtree. This only requires a constant number of
pointer changes in the quadtree and can be done in $O(1)$ worst-case time, and we
may also need to restore balance to the quadtree, which requires $O(1)$ worst-case
time by Lemma~\ref {lem:quadtree-balance}.  Finally, we follow
pointers from the quadtree to the query tree, and perform the corresponding
deletion and insertion in that tree, which by
Lemmas~\ref{lem:eot-dynamic} and~\ref{lem:eot-quadtree} takes
only constant time. Thus, the entire update can be completed in
$O(\log \rho)$ worst-case time. 
  
Note that we can also insert or delete intervals from the data structure in
$O(\log n)$ time; we perform a query to locate where the interval belongs and a
local update to insert it or remove it. 
\begin{theorem}
We can maintain a linear size data structure over a set of $n$
non-overlapping intervals such that we can perform point location queries and
insertion and deletion of intervals in
$O(\log n)$ worst-case time and local updates in
$O(\log \rho)$ worst-case time.
  \end{theorem}
\fi

\section {Lower Bounds} \label {sec:lower}

  In this section, we will investigate lower bounds on updates. Clearly, there cannot be any non-trivial lower bounds if we do not restrict the time we allow to spend on queries, so we will restrict our attention to data structures that support $O (\log n)$ queries.
  We will first argue that insertions must take $\Omega (\log n)$ time, and then extend the argument to show that updates cannot be implemented any faster unless they are local. Finally, we show that some of the restricted settings we use are necessary.

\subsection {Insertions and deletions}

The relationship between preprocessing time, insertion time, and query time in dynamic data structures is well-studied. Borodin \etal~\cite {bgly-esupo-81} first showed that if membership queries in an ordered set need to be supported in sublinear time, then insertions must necessarily take $\Omega (\log n)$ comparisons.
The seminal paper by Ben-Or~\cite{DBLP:conf/stoc/Ben-Or83}, relating the height of a computation tree to the connected components in the space of possible inputs to a problem, made it possible to make the same argument in algebraic computation trees.
We base our lower bounds on a reduction to the semi-dynamic
membership problem, which was shown by Brodal and Jacob~\cite{DBLP:conf/focs/BrodalJ02} to have a
$\Omega(\log n)$ lower bound for queries and insertions on a Real RAM.


\begin {theorem} [from \cite{DBLP:conf/focs/BrodalJ02}]
  Let $\mathcal D$ be a data structure that maintains a set $S$ of $n$ real numbers that supports insertions
  in $I(n)$ time and membership queries in $Q(n)$ time.
  Then we have $I(n) = \Omega \left(\log {\frac n {Q(n)}} \right)$.
\end {theorem}

From this result, we easily obtain a $\Omega (\log n)$ lower bound on insertions for our problem.

\begin {corollary}
  Let $D$ be a data structure that stores a set $\mathcal R$ of $n$ regions in
  $\R^d$, and allows for point location queries in $Q(n)$ time and
  insertions/deletions in $I(n)$ time.
  If $Q(n) = o(n)$, then $I(n) = \Omega (\log n)$.
\end {corollary}


\subsection {Local updates}

To obtain lower bounds on the complexity of local updates, the standard approach does not
work directly.
After all, every element that gets moved locally must have been inserted before, so in any
static argument involving $n$ elements we already need to spend $\Omega(n \log n)$ time
just to initialize the structure.
Instead, we will argue that when the local updates are sufficiently powerful, we may start with a data structure that already contains $n$ elements, and use the local updates to simulate insertions. 
If we identify an invariant on the elements and show that it is maintained after the updates, we can simulate arbitrarily many rounds of insertions, and their processing time can no longer be charged to the initial (true) insertions into our data structure.

\begin {lemma} \label{lem:lower-bound1}
  Let $D$ be a data structure that stores a set $\mathcal R$ of $n$ regions in
  $\R^d$, and allows for point location queries in $Q(n)$ time and updates in
  $U(n)$ time. Let $\mathcal R$ be a set of regions on which there exists some order
  $\mathcal O : \mathcal R \to \N$.
  Suppose that for any
  permutation $\pi$ of $n$ elements, there exists a sequence of $O(n)$ updates
  $S_\pi$ that turns $\mathcal R$ into $\mathcal R'$ such that 
  $\mathcal O(\mathcal R) = \pi(\mathcal O (\mathcal R'))$.
   Then if $Q(n) = o(n)$,
  $U(n) = \Omega (\log n)$.
\end {lemma}
  
The above lemma is a fairly straightforward consequence of \cite{bgly-esupo-81}.
\maarten {Is it? I have trouble reconstructing the straightforward argument...}

\joe{removed alternate paragraph}

\paragraph {Unbounded moving.}

We first show that if we only allow to move regions (not scale them), but have no bound on the distance they may move, we still have a $\Omega (\log n)$ lower bound. 

\begin {lemma} \label {lem:lowerbound-moves}
  Let $D$ be a data structure that stores a set $\mathcal R$ of $n$ disjoint regions in
  $\R^d$, and allows for point location queries in $Q(n)$ time,
  insertions in $I(n)$ time,
  and move updates in $U(n)$ time. 
  If $Q(n) = o(n)$, then $U(n) = \Omega (\log n)$.
\end {lemma}
\begin {proof}
  Let $\mathcal I$ be the set of intervals $\{I_i = [i, i+1) \mid i = 1, \ldots, n\}$, and let $\mathcal I_j$ be $\mathcal I$ translated by $jn$.
  Given any permutation $\pi$ on $n$ elements, there is clearly a sequence of $n$ move
  updates that takes the elements of $\mathcal I_j$ and turns them into 
  $\pi(\mathcal I_{j+1})$: every element can move directly to its new location. 
  Therefore, by Lemma~\ref {lem:lower-bound1}, we must have $U(n) = \Omega (\log n)$.
  Since all intervals of $\mathcal I_j \cup \mathcal I_{j+1}$ are disjoint, no interval will overlap any other interval during the execution of the updates, and we maintain a ply of $1$.
\end {proof}

\paragraph {Unbounded scaling.}

  If we restrict moving distances but allow full freedom in the precision changes,
  and allow the ply to become $2$, then the above argument can be trivially adapted:
  We can permute the set of intervals by first grow each interval large enough to contain the whole domain of
  interest, and then shrink it to its new location. 
  We now show that even if we insist on disjoint intervals, there is still a $\Omega (\log n)$ lower bound on
  unrestricted scaling, even if we only either grow or shrink the intervals.

\begin{lemma}\label{lem:lower-bound2}
Let $D$ be a data structure that stores a set $\mathcal I$ of $n$ intervals in
$\R^1$ and allows for point location queries in $Q(n)$ time and updates in
$U(n)$ time subject to the following restrictions:
No more than
one interval is allowed to overlap any point;  
An update may replace interval $I$ by interval $I'$ that is within distance
$2 |I|$, and
has size $0 < |I'| \leq 2 |I|$ (i.e. the interval can shrink arbitrarily).
Then, if $Q(n) = o(n)$, $U(n) = \Omega (\log n)$.
\end{lemma}
\begin{proof}
Let $\mathcal I = \{I_i = [2^i, 2^{i+1}) \mid i = 1, \ldots, n\}$. That is,
the intervals have their left endpoints aligned on powers of 2, have
exponentially increasing size, and do not intersect each other.
Let $\mathcal I_j$ be $\mathcal I$, scaled down by a factor $2^{jn}$.
Then in a local update any interval from $\mathcal I_j$ can be mapped to any interval in $\mathcal I_{j+1}$.
 Therefore, in $n$ updates which never cause the
ply to exceed one, the order of the intervals can be permuted arbitrarily. 
Thus, by Lemma~\ref{lem:lower-bound1} if $Q(n) = o(n)$, then $U(n) = \Omega (\log n)$.
\end{proof}

Note that by reversing the direction of the updates, the same argument holds for arbitrary growing without shrinking.

%


\paragraph {Unbounded skinniness.}

When $d > 1$, we additionally require the regions to be fat. 
Without this requirement, it is not obvious how one should define similarity of regions.
Using the definition from Section~\ref {sec:prob}, we can easily adapt the above interval constructions to skinny rectangles.

\begin {lemma}
  Let $D$ be a data structure that stores a set $\mathcal R$ of $n$ disjoint regions in
  $\R^d$, $d \ge 2$, and allows for point location queries in $Q(n)$ time,
  insertions in $I(n)$ time,
  and similar updates in $U(n)$ time. 
  If $Q(n) = o(n)$, then $U(n) = \Omega (\log n)$.
\end {lemma}
\begin {proof}
  Let $\mathcal I$ be the set of intervals as constructed in Lemma~\ref {lem:lowerbound-moves}.
  Extend the intervals to a set of rectangles $\mathcal R = \mathcal I \times [0,n]$. 
  Now all elements in $\mathcal R$ have diameter bigger than $n$.
  Every update in the proof of Lemma~\ref {lem:lowerbound-moves} moves an interval over a distance of at most $2n$; clearly, the corresponding update of the rectangle is $3$-similar.
\end {proof}

On the other hand, if all regions are convex and homothetic as in the proof
above, then they can be made fat by simply scaling the plane. It would be
interesting to investigate alternative, more restrictive definitions of
similarity that capture this effect, and analyze to what extent 
On the other hand, if all regions are convex and homothetic as in the proof above, then they can be made fat by simply scaling the plane. It would be interesting to investigate alternative, more restrictive definitions of similarity that capture this effect, and analyze to what extend local updates on non-fat objects can then be supported.

\ifFull
\joe{we have an ifFull flag on this paragraph, even though it's in appendix?}
\maarten{I think last time we didn't want to include this paragraph in the submission because it's a little weak, but we did want to keep it in the full version so we wouldn't forget about it and hopefully in a moment of sudden insight figure out what to do with it. I don't think anything changed since last time. Similar for the fairly straightforward consequence in the previous section.}
\paragraph {Unbounded ply.}
If we allow the regions to overlap arbitrarily, then clearly a single update can cause a
linear number of changes to a subdivision in the plane. Thus, no method which
explicitly maintains the regions will be able to handle such updates. Moreover, it seems
impossible to maintain the set of regions implicitly without requiring some sort of
hierarchical subdivision of the regions, which would then require updates to take at least
$\Omega(\log n)$ time.

\joe{omitted alternate ply paragraph}
\ifFull
Similarly, the update complexity may depend on the current 
\emph{ply} of the regions (that is, the number of regions which intersect a common
sub-region.) If all regions contain a common
interior, then in $O(n)$ shrink operations we can permute them arbitrarily
within the common region,
which again implies a $\Omega(\log n)$ lower bound.
However, even if we restrict the ply, and even in the one-dimensional case, 
we have a $\Omega(\log n)$ lower bound for updates if we
want to allow arbitrary shrinking. 
\joe{the old paragraph isn't really about restrictions on the ply, especially when we include
the last sentence. I took a stab at a new paragraph, but we may want to omit it.}
\maarten {Looks better. But still I think I'm in favour of omission unless we have something more interesting to say...}
\fi

\fi

\section {Extensions} \label {sec:extensions}

We show how to extend our data structure so that
\begin{itemize}
  \item Our compressed quadtree does not require the \emph {floor} operation.
  \item The marked-ancestor component can be implemented
  on a pointer machine.
\end{itemize}

  \subsection {Arbitrary scales and compressed quadtrees}
  \label {sec:ext-nofloor}
  
    In this paper, we assumed that compressed nodes in a quadtree are aligned with their parents. However, aligning a node at an arbitrary scale is not supported in constant time on a Real RAM, unless we can use the \emph {floor} operation (or a different non-standard operation~\cite[Chapter~2]{HarPeled11}). While this is a very natural assumption in practice and does not hinder the implementation of our algorithms, it also is ``unreasonably powerful'' in theory, so we would like to avoid its use to strengthen our theoretical bounds.
    
    A standard way to avoid this problem in the literature is to allow compressed nodes to be associated with \emph {any} square that is contained in the parent square and sufficiently small~%
    \cite {blmm-pipstse-10,HarPeled11,lm-qtdte-11}.
    This is fine in a static context, but in our dynamic quadtrees we have to be more careful: after a number of merge operations the size difference between a compressed node and its parent may become less than a factor $a$, and then we cannot simply connect the two trees since they are not aligned.
    
However, in a compressed quadtree with non-aligned compressed nodes we can still
 align nodes when necessary in $O(1)$ amortized time, which we now show.
 We will view each compressed node as a cut, which divides its ancestors and
 descendants into different components which may not be aligned with each other.
 Let $n$ be the number of nodes in the quadtree and let $N_i$ be the number of
 nodes in component $i$.
 We define our potential function for each component as
 \[
 \Phi_i = N_i (\log n - \log N_i)
 \]     
 and the total potential function as $\sum_i \Phi_i$.

We now analyze the cost of local or global update operation. 
\begin{itemize}
\item \emph{insert into existing component:} \\
The insertion takes $O(\log n)$, and adds $O(1)$ nodes to this component of the quadtree. 
For each node we add, we increment $n$ and $N$ by 1. 
Therefore the change in potential of the component is
\begin{align*}
\Delta \Phi  &= - N_i (\log n - \log N_i) + (N_i + 1)(\log(n + 1) - \log(N_i + 1))\\
 & = N_i (\log \frac{n+1}{n} - \log \frac{N_i+1}{N_i}) + (\log (n + 1) - \log (N_i + 1))
 \\
 & = \log (n + 1) + \mbox{ negative terms} \\
 & = O( \log n)
\end{align*}
Therefore, the total amortized cost is $O(\log n + \Delta \Phi)= O(\log n)$.

\item \emph{insert into new component:} \\
An insertion may create a new compressed node. In this case, we create the
corresponding component, and for each of the $O(1)$ nodes created in the new
component, we  have the following change in potential:
\begin{align*}
\Delta \Phi &= 1 \cdot (\log n - \log 1) = O(\log n)
\end{align*}
Therefore the total amortized cost in this case is also $O( \log n)$.

\item \emph{merge 2 components:} \\
If the size difference between the two components becomes less than a factor of
$a$, then they must be merged. We must make sure that the two components are
aligned, and so we spend $O(1)$ time for each node in the smaller component to
align them with the larger component. Let $N_s$ be the number of nodes in the
smaller component, $N_L$ be the number of nodes in the larger component and 
$N = N_s + N_L$ be the total number of nodes in both components. Note
that $N \geq 2 N_s$.  
The change in potential for these two components is
\begin{align*}
\Delta \Phi & =  - N_s(\log n - \log N_s) - N_L(\log n - \log  N_L) + N(\log n - \log N) \\
        & = N_s \log N_s + N_L \log N_L - N \log N \\
      & = N_s (\log N_s - \log N) + N_L (\log N_L - \log N) \\
      & < N_s (-\log 2) + N_L (-\log \frac{N}{N_L}) \\
      & < - N_s
\end{align*}
Therefore the amortized cost of merging the two components is $O(N_s - N_s) =
O(0)$.

\item \emph{deletion:} \\
If we delete a node out of a component containing $N$ nodes, then  the change in potential is
\begin{align*}
\Delta \Phi & = -N (\log n - \log N) + (N - 1) (\log (n - 1) - \log (N - 1)) \\
& = N \log \frac{N}{N-1} - N \log \frac{n}{n-1} - (\log (n - 1) - \log (N - 1)) \\
& < O(1)
\end{align*}
Therefore the amortized cost of a deletion is $O(\log n + 1) = O(\log n)$.
\item \emph{local updates:} \\
Local updates do not change the total number of nodes, and move at most $O(1)$
nodes from 1 component to another. Therefore, the change in potential of a local
update is $O(1)$, and the amortized cost is $O( \log \log n + 1 ) = O(\log \log
n)$.
\end{itemize}
\joe{reviewer 1 complains the following lemma is not clearly proved}
   \begin{lemma}
  We can align compressed subtrees by the time they are connected in 
  $O(1)$ amortized time per split or merge operation. 
   \end{lemma}
   
\subsection{Marked-ancestor queries on a pointer machine}
\label{sec:MAT}
We now show how to adapt the marked-ancestor structure of 
Alstrup~\etal{}~%
    \cite{ahr-map-98,ahr-map-98-conf}
so that it works on a pointer machine.
 
Suppose that we are given a tree $T$ over which we want to support marked
ancestor queries. Recall that a \emph{heavy node} is a node with at least two
children.
Alstrup~\etal{} maintain what they call an ART-universe. That
is, they partition the nodes of $T$ into micro-trees such that each micro
tree has at most $O(\log n)$ heavy nodes, and any leaf to root path passes
through at most $O(\log n / \log \log n)$ micro trees. 
Thus, they reduce any marked-ancestor query in $T$ to at most $O(\log n / \log
\log n)$ \emph{exists} queries on the micro trees which determine if each micro
tree on the path to the root 
contains a marked ancestor and one marked-ancestor query in the first micro-tree
which contains a marked ancestor. The final marked-ancestor query in the micro
tree is answered by determining which of the at most $O(\log n)$ paths in the
micro tree contains a marked ancestor, and then performing a marked successor
query on that path. 

The reduction from queries in $T$ to queries in micro-trees only requires a
pointer machine.
However, they require a word-RAM to support
their queries within micro-trees in two places. First, they maintain
connectivity between the $O(\log n)$ paths within a micro-tree 
using the bit-manipulation techniques of \cite{ass-oodct-97}. 
Second, they use a RAM based implementation to support their marked successor
queries on the marked path.   
Thus, if we replace these two data structures, we will support all operations on
a pointer machine. 

The latter data structure is easy to replace. We just use a pointer-machine
based implementation of a Union-Split-Find data structure 
    \cite{k-aru-84-tr,mehlhorn-up,mn-dfc-86-tr,mna-lbcusfp-88}
to support the marked successor queries on a path. We now describe how to
replace the former data structure.

We keep the same subdivision of a micro-tree into $O(\log n)$ paths, but instead
of using bit-manipulations to keep track of the $O(\log n)$ paths, we build a tree on the
paths. By construction, each path does not contain any heavy nodes in its
interior. Therefore, we can compress each path in the micro-tree to a single
node representing the path, where each compressed-path-node is marked if and only if at
least one node on the corresponding path is marked. 
The result is a tree with a logarithmic number
of nodes. Over our path-node-tree, we build the Link-Cut data structure of
Sleater and Tarjan~\cite{st-dsdt-83}, which maintains a dynamic forest and supports 
operations link, cut, and find-root in $O(\log N)$ time, where
$N$ is the number of nodes in the forest. 
Just as Union-Split-Find is equivalent to the marked successor problem, the
link-cut trees support all the operations required for the marked-ancestor
problem. 
The Link operation corresponds to
unmark, and the Cut operation corresponds to the mark operation. Likewise, the
find-root operation, which returns the root of the current tree corresponds to
the marked-ancestor query. 
Since the number of nodes in our path-node-tree is $N = O(\log n)$, this data structure
supports all marked-ancestor operations on the path-node-tree in $O(\log \log
n)$ time.

Thus, all of the components of the data structure are now supported on a pointer
machine.
To perform a query in $T$, we perform at most $O(\log n / \log \log n)$ marked
ancestor queries in the micro-trees. When we reach the first marked path-node in
a micro-tree, we also perform a marked successor query on this path, and the
returned node is the
first marked ancestor in $T$. 
Since the time spent in each micro-tree is at most $O(\log \log n)$, the total
time required for a query in $O(\log n)$.

To perform a mark/unmark update of a node $v \in T$, we perform the
corresponding update on the path $P$ containing $v$. If this is/was the only marked
node in $P$, then we also update the corresponding path node $u_P$ in the link-cut
data structure containing $u_P$. Thus we update a constant number of data
structures, and each update takes $O(\log \log n)$ time. 
\begin{lemma}
We can maintain a data structure over any rooted tree $T$ which supports
insertions and deletions of leaves in $O(1)$ amortized time, marking and
unmarking nodes in $O(\log \log n)$ worst-case time, and queries for the lowest marked
ancestor in $O(\log n)$ worst-case time.
All operations are supported on a pointer machine.
\end{lemma}

\end {document}